\documentclass[aps,prl,twocolumn,nofootinbib,superscriptaddress]{revtex4-2}

\pdfoutput=1

\usepackage[utf8]{inputenc} 
\usepackage{amsmath}
\usepackage{amssymb}
\usepackage{xcolor}
\usepackage{graphicx}
\usepackage{dcolumn}
\usepackage{bm}
\usepackage{microtype}
\usepackage{threeparttable}
\usepackage{mathtools}
\usepackage[colorlinks=true]{hyperref}
\usepackage{physics}
\usepackage{comment}

\newcommand{\be}{\begin{eqnarray}}
\newcommand{\ee}{\end{eqnarray}}

\begin{document}

\title{Geometric Interpretation of Timelike Entanglement Entropy}

\author{Michal P. Heller}
\email{michal.p.heller@ugent.be}
\affiliation{Department of Physics and Astronomy, Ghent University, 9000 Ghent, Belgium}

\author{Fabio Ori}
\email{fabio.ori@ugent.be}
\affiliation{Department of Physics and Astronomy, Ghent University, 9000 Ghent, Belgium}

\author{Alexandre Serantes}
\email{alexandre.serantesrubianes@ugent.be}
\affiliation{Department of Physics and Astronomy, Ghent University, 9000 Ghent, Belgium}

\begin{abstract}

\noindent Analytic continuations of holographic entanglement entropy in which the boundary subregion extends along a timelike direction have brought a promise of a novel, time-centric probe of the emergence of spacetime. We propose that the bulk carriers of this holographic timelike entanglement entropy are boundary-anchored extremal surfaces probing analytic continuation of holographic spacetimes into complex coordinates. This proposal not only provides a geometric interpretation of all the known cases obtained by direct analytic continuation of closed-form expressions of holographic entanglement entropy of a strip subregion but crucially also opens a window to study holographic timelike entanglement entropy in full generality. We initialize the investigation of complex extremal surfaces anchored on a timelike strip at the boundary of anti-de Sitter black branes. We find multiple complex extremal surfaces and discuss possible principles singling out the physical contribution.
\end{abstract}

\maketitle

\begin{figure}[b]
    \centering
    \includegraphics[width=0.9\linewidth]{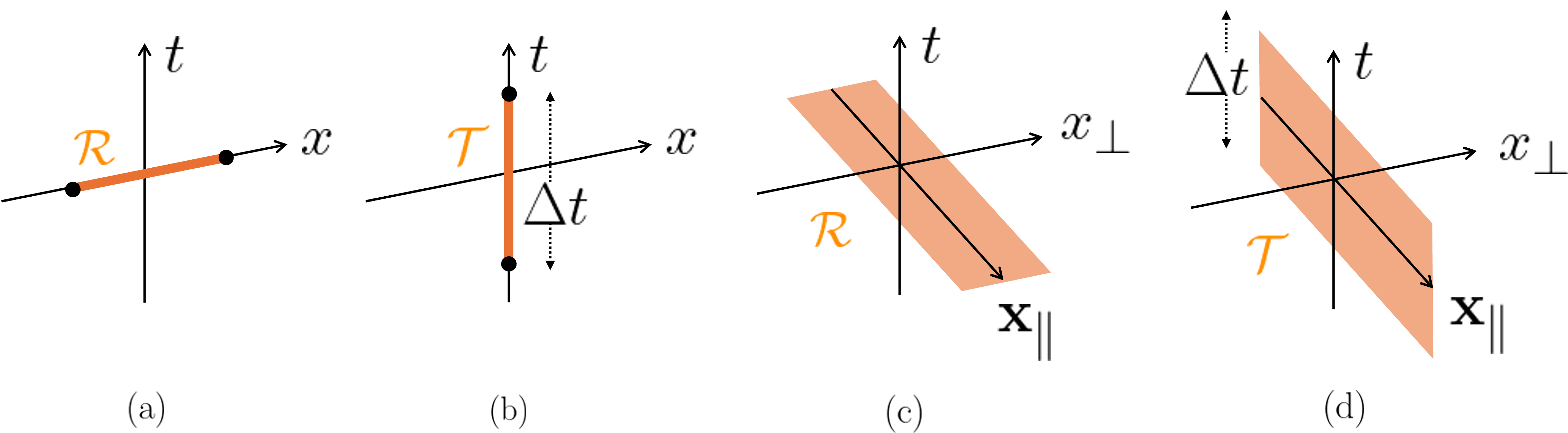}
    \caption{(a) Spatial single interval subregion in 1+1 dimensions and (c) its higher-dimensional generalization as a strip. (b,d): Analogous timelike regions can be obtained from (a,c) by making one of the spatial coordinates imaginary.}
    \label{fig:regions}
\end{figure}

\noindent \textbf{\emph{Introduction and summary.--}} Entanglement entropy (EE) has proven to be a prolific notion across the contemporary physics landscape~\cite{Faulkner:2022mlp}. In the spacetime picture of quantum mechanics [see Figs.~\ref{fig:regions}(a) and \ref{fig:regions}(c)], it is defined by picking a time slice that gives rise to a state and considering a spatial subregion on this time slice giving rise to a reduced density matrix. EE is then defined as the von Neumann entropy of this reduced density matrix. While in general very hard to compute, remarkably the EE for spatial bipartitions acquires a simple geometrical description in strongly coupled quantum field theories with many microscopic constituents, where its holographic dual are extremal codimension-two surfaces anchored at the asymptotic boundary on the edge of the relevant spatial subregion~\cite{Ryu:2006bv,Hubeny:2007xt,Casini:2011kv,Lewkowycz:2013nqa,Dong:2016hjy}. The holographic EE (HEE) is subject to the conditions of homology \cite{Headrick:2007km} and minimality to pick the relevant extremal surface if multiple ones exist.

Recently, Refs.~\cite{Doi:2022iyj,Doi:2023zaf} pursued a brilliant idea to depart from the standard definition of EE and instead consider an analog problem in which the subregion extends also in a timelike direction at the expense of a spacelike one.
In two-dimensional conformal field theories (CFTs$_{2}$) a paradigmatic example of a subregion is a single spatial interval and the idea then is to consider a single timelike interval [see Fig.~\ref{fig:regions}(b)]. Subsequently, Refs.~\cite{Doi:2022iyj,Doi:2023zaf} considered several known closed-form expressions for EE---the universal CFT$_{2}$ prediction for a single interval and the HEE for a strip subregion in the vacuum---as a functional of parameters specifying the boundary subregion, and performed an analytic continuation to make the extent of the subregion timelike [see Figs.~\ref{fig:regions}(b) and ~\ref{fig:regions}(d)]. This analytic continuation indicated that the quantity obtained this way, dubbed holographic timelike EE (HTEE), is a complex-valued pseudoentropy~\cite{Doi:2022iyj,Doi:2023zaf}.

In three-dimensional holography, it was possible for Refs.~\cite{Doi:2022iyj,Doi:2023zaf} to identify candidate, partly spacelike and partly timelike, bulk geodesics whose respective real and imaginary lengths reproduce the analytic continuation of the EE of a single subregion. 
Unfortunately, beyond these cases, no geometric picture exists for what the HTEE could be and no prescription exists to calculate it for general timelike-extended subregions. We believe this is an important problem to alleviate. One reason is the connection between holography and tensor networks~\cite{Swingle:2009bg}, with the latter community considering closely related quantities in the context of unitary time evolution under the umbrella of temporal entanglement~\cite{Banuls:2009jmn,Hastings:2014qqa,sonner2021influence,lerose2021scaling,giudice2022temporal,lerose2023overcoming,Carignano:2023xbz,Carignano:2024jxb}. Another is that the HEE and other geometric probes of the emergent spacetime, including correlators of heavy operators~\cite{Balasubramanian:1999zv}, Wilson loops~\cite{Maldacena:1998im,Rey:1998ik} and holographic complexity~\cite{Stanford:2014jda,Brown:2015bva,Couch:2016exn,Belin:2021bga}, have their limitations, e.g.\ when it comes to probing black hole interiors~\cite{Fidkowski:2003nf,Engelhardt:2013tra}, and it is important to look for probes with complementary virtues. Finally, accelerated expansion often rules out standard extremal hypersurfaces in de Sitter universes~\cite{Susskind:2021esx,Chapman:2021eyy,Aalsma:2022eru,Chapman:2022mqd,Doi:2022iyj,Doi:2023zaf,Narayan:2022afv,Narayan:2023zen}. 

\begin{figure}[h!]
    \centering
    \includegraphics[width=0.8\linewidth]{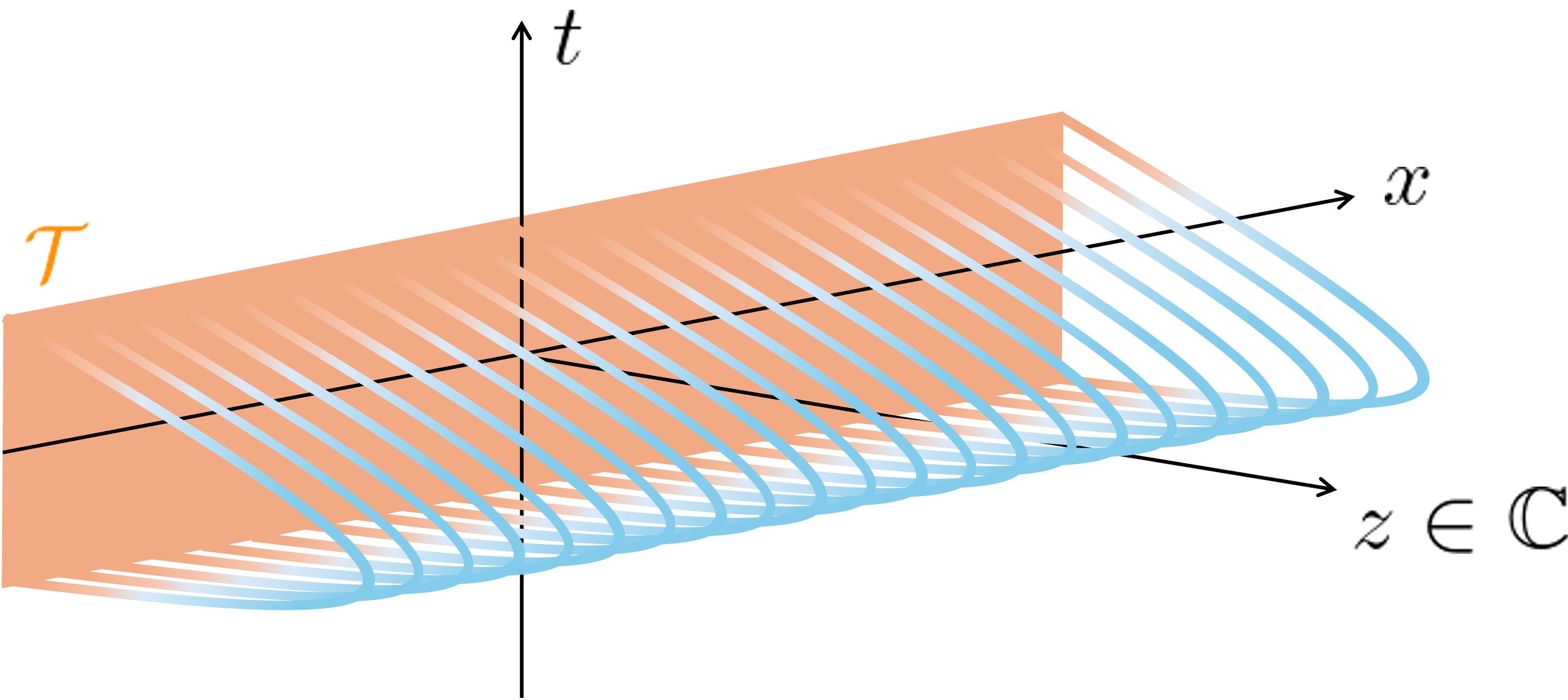}
    \caption{Our HTEE proposal~\eqref{proposal} entails considering codimension-two complex (blue) extremal surfaces that are anchored on the asymptotic boundary on a desired real (red) timelike subregion, here the timelike strip from Fig.~\ref{fig:regions}(d).}
    \label{fig:prescription}
\end{figure}

\begin{center}
\setlength{\fboxsep}{.02\columnwidth}
\setlength{\fboxrule}{.01\columnwidth}
\fcolorbox{gray!60}{gray!20}{%
    \parbox{.93\columnwidth}{%
We propose that the bulk carriers of  HTEE are codimension-two extremal surfaces~$\gamma_\mathcal{T}$ anchored on a timelike boundary subregion~$\mathcal{T}$ and in general extending in a complexified bulk geometry, see Fig.~\ref{fig:prescription}. The HTEE is then proportional to the area of $\gamma_\mathcal{T}$,
\begin{equation}\label{proposal}
S_\mathcal{T} = \frac{A(\gamma_\mathcal{T})}{4G},
\end{equation}
where $G$ is the bulk gravitational constant and the normalization reproduces HEE upon analytic continuation. 
With the hindsight of examples, in the outlook we discuss various physical conditions to select among possible multiple contributing extremal surfaces. 
}}\end{center}

By a complexified geometry we mean a holographic geometry in which coordinates become complex variables with the asymptotic boundary being defined as a real locus in the standard way. This naturally connects to earlier studies of complex geodesics in holography in the context of black hole singularity~\cite{Fidkowski:2003nf} and correlators of heavy operators at timelike separations~\cite{Balasubramanian:2012tu}. Through~Eq.~\eqref{proposal}, HTEE is explicitly a geometric object and can in principle be determined for a timelike subregion of any shape in any state. We view our proposal as a conservative generalization of the basic building block of the HEE prescription to timelike boundary subregions, which allows to utilize various techniques and concepts from the study of HEE.

We check that our proposal reproduces all the known cases of HTEE obtainable via analytic continuation. However, our geometric interpretation departs from the one provided by Refs.~\cite{Doi:2022iyj,Doi:2023zaf} in the context of three-dimensional holography. There, the real part of the HTEE came from spatial geodesic segments and the imaginary part from timelike geodesic segments in the same spacetime. In our case, both parts are generically geometrically inseparable and originate from geodesics probing bulk spacetime coordinates having both real and imaginary parts across the relevant curve. Within our proposal, the interpretation in terms of a combination of timelike and spatial paths is scarce and typically subtle (see Supplemental Material).

Finally, to demonstrate the predictive power of our proposal, we determine the HTEE for a timelike strip on the boundary of a black brane geometry. This example connects with the notion of critical surfaces underlying the tsunami picture of EE production in holographic quenches~\cite{Hartman:2013qma,Liu:2013iza,Liu:2013qca} and, crucially, gives rise to multiple complex extremal surfaces satisfying the same boundary conditions. We discuss two possible criteria---minimality of real part of the area and consistency with the ultraviolet-infrared (UV--IR) correspondence~\cite{Susskind:1998dq}---that could select the physical contribution. 

\vspace{10 pt}

\noindent \textbf{\emph{Setup}.--} The strip subregion of interest, depicted in Fig.~\ref{fig:regions}(d), is living in $d$-dimensional Minkowski spacetime located on the regularized ($z  = \epsilon \ll 1$) boundary~of
\begin{equation}\label{metric}
ds^2 = \frac{1}{z^2}\left(-f(z) dt^2 + \frac{dz^2}{f(z)} + d\textbf{x}^2 \right),        
\end{equation}
where the curvature scale is unity. The choice $f(z) = 1$ corresponds to the empty anti-de Sitter (AdS) space encapsulating the vacuum of the dual CFT, whereas $f(z) = 1 - \frac{z^d}{z_H^d}$ corresponds to a black brane encapsulating a thermal state. The strip is defined as ($\textbf{x} = \{\textbf{x}_\parallel, x_\perp\}$)
\begin{equation}\label{strip}
\mathcal{T} \equiv \left\{(t,\textbf{x}): t\in \left[-\frac{\Delta t}{2},\frac{\Delta t}{2}\right],\textbf{x}_\parallel \in \mathbb{R}^{d{-}2},x_\perp{=}0 \right\},
\end{equation}
and acts as a timelike entangling region on the asymptotic boundary that anchors the extremal surface~\eqref{proposal}.

By symmetry, the codimension-two bulk extremal surface $\gamma_T$ takes the form 
\begin{equation}\label{extremal_surface}
X^\mu(\lambda) = \{t_s(\lambda),z_s(\lambda),\textbf{x}_\parallel, x_\perp = 0\},     
\end{equation}
where $\lambda$ is a parameter moving along the variable part of the surface. Given this, according to our proposal \eqref{proposal} we need to extremize the area density functional, 
\begin{equation}\label{action}
\hspace{-5 pt} \mathcal{A} {\equiv} \frac{A}{V} {\equiv} \int d\lambda \mathcal{L} {\equiv} \int d\lambda \sqrt{\frac{-f(z_s(\lambda))t_s'(\lambda)^2 +\frac{z_s'(\lambda)^2}{f(z_s(\lambda))}}{z_s(\lambda)^{2d-2}}},   
\end{equation}
to find the HTEE density $\mathcal{S}_\mathcal{T} \equiv S_\mathcal{T}/V$. In these expressions, $V$ stands for the volume of $\mathbb{R}^{d-2}$ spanned by $\textbf{x}_\parallel$. Note that for $d = 2$ there are no $\textbf{x}_\parallel$ directions and~$V \equiv 1$. 

Since the bulk metric \eqref{metric} does not depend on time, $\gamma_\mathcal{T}$ has an associated conserved quantity, $p$, such that the Euler--Lagrange equations stemming from \eqref{action} can be reduced to the first-order form
\begin{equation}\label{t_eq}
t_s'(\lambda)^2 = \frac{p^2 z_s(\lambda)^{2d-2} z_s'(\lambda)^2}{f(z_s(\lambda))^2(f(z_s(\lambda)) + p^2 z_s(\lambda)^{2d-2})}.     
\end{equation}
From Eq.~\eqref{t_eq} it is immediate to see that the locus $z=z_t$ where $f(z_t) + p^2 z_t^{2d-2} = 0$ corresponds to a tip of $\gamma_\mathcal{T}$ where $t_s(z)$ has a branch-point singularity.  See also Fig.~\ref{fig:prescription}.
\\\\
\noindent \textbf{\emph{Crosschecks.--}} We will show now that the proposal \eqref{proposal} reproduces the HTEE in cases where it can be computed explicitly via analytic continuation of areas of HEE extremal surfaces~\cite{Doi:2022iyj,Doi:2023zaf}. In these cases, the proposal can be thought of as a direct analytic continuation of the surfaces themselves, rather than of their areas alone.

\vspace{5pt}

\noindent \underline{\emph{AdS$_3$ holography}.} In this case $\gamma_T$ will be a boundary-anchored bulk geodesic. We choose $\lambda$ as an affine parameter, $\mathcal{L} = 1$, such that the end points of the bulk geodesic at the asymptotic boundary are reached at $\lambda = \mp \lambda_\star$, 
\begin{equation}\label{ads3_bc}
z_s(\pm \lambda_\star) = \epsilon, \quad t_s(\pm \lambda_\star) = \pm \Delta t/2.
\end{equation}
For the vacuum state, the solution of Eq.~\eqref{t_eq} subject to the boundary conditions \eqref{ads3_bc} is given by
\begin{subequations}
\begin{equation}\label{ads3_sol1}
\hspace{-8 pt} t_s(\lambda) = \sqrt{\frac{\Delta t^2}{4}-\epsilon^2}\tanh(\lambda), \quad 
z_s(\lambda) = \frac{i \sqrt{\frac{\Delta t^2}{4} - \epsilon^2}}{ \cosh(\lambda)},  
\end{equation}
\begin{equation}\label{ads3_sol2}
\lambda_\star = \log \frac{\Delta t}{\epsilon} + i\frac{\pi}{2} + O(\epsilon^2).
\end{equation}
\end{subequations}
The regularized geodesic length, $L = 2\,\lambda_\star$, reproduces the HTEE of a timelike segment in the vacuum state of a Minkowski space CFT$_{2}$ with central charge $c$~\cite{Doi:2023zaf}  
\begin{equation}\label{ads3_vacuum_ent}
S_\mathcal{T} = \frac{L}{4 G} = \frac{c}{3} \log \frac{\Delta t}{\epsilon} + i \frac{\pi c}{6}\,,  
\end{equation}
where we used $c \equiv 3/(2G)$~\cite{Brown:1986nw}.

For a black brane, the solution of Eq.~\eqref{t_eq} with the boundary conditions \eqref{ads3_bc} is given by
\begin{subequations}\label{ads3_thermal_sol}
\begin{equation}
\hspace{-10 pt}\frac{t_s(\lambda)}{z_H}{=}\frac{1}{2}\log \frac{\cosh(\lambda{+}\frac{\Delta t}{2z_H})}{\cosh(\lambda{-}\frac{\Delta t}{2z_H})},~ 
\frac{z_s(\lambda)}{z_H}{=}\frac{i\sinh(\frac{\Delta t}{2 z_H})}{\cosh(\lambda)},  
\end{equation}
\begin{equation}
\lambda_\star = \log \frac{2 z_H \sinh(\frac{\Delta t}{2z_H})}{\epsilon} + \frac{i \pi}{2}, 
\end{equation}
\end{subequations}
where we have quoted the expressions at leading order in~$\epsilon$. The proposal \eqref{proposal} gives then 
\begin{equation}
\label{eq.HTEEbh3}
S_\mathcal{T} = \frac{L}{4 G_3} = \frac{c}{3} \log \frac{2 z_H \sinh(\frac{\Delta t}{2 z_H})}{\epsilon} + \frac{i \pi c}{6},    
\end{equation}
which also agrees with the findings of Ref.~\cite{Doi:2023zaf}. 

Besides demonstrating that the proposal~\eqref{proposal} computes correctly the HTEE in known cases, these two simple computations also illustrate a crucial aspect of it: namely, since  $\lambda_\star$ is \emph{complex}, the bulk geodesic $X^\mu(\lambda)$ has to be thought of as a 3-tuple of complex functions of a complex affine parameter $\lambda$. From this perspective, any path in the complex $\lambda$-plane joining $-\lambda_\star$ and $\lambda_\star$ provides a valid section of the complex geodesic. Among this infinite-dimensional set, there happen to exist special paths singled out by their reality properties, which allow for a direct comparison with the geometric interpretation of the HTEE proposed in Ref.~\cite{Doi:2023zaf} (see Supplemental Material).

\vspace{5pt}

\noindent \underline{\emph{Higher-dimensional holography.}} In $d>2$, the HTEE for~\eqref{strip} is only known in the vacuum~\cite{Doi:2023zaf},
\begin{small}
\begin{equation}\label{ST_vacuum_higher-d}
\hspace{-8 pt}\mathcal{S}_\mathcal{T} = \frac{\left(\frac{1}{\epsilon^{d-2}} + \frac{c_d}{2} \frac{(-i)^d}{\Delta t^{d-2}}\right)}{2 (d{-}2)G}
,~~
c_d = \left(\frac{2 \sqrt{\pi}\,\Gamma\left(\frac{d}{2(d-1)}\right)}{\Gamma\left(\frac{1}{2(d-1)}\right)}\right)^{d{-}1}.    
\end{equation}
\end{small}
This result follows from the analytic continuation of the HEE of a spacelike strip from real to imaginary width. Here we demonstrate that the HTEE proposal \eqref{proposal} naturally reproduces Eq.~\eqref{ST_vacuum_higher-d} and provides for the first time a clear a geometrical understanding of this result.   

To perform the computation, it is convenient to employ diffeomorphism invariance to set $z_s(\lambda) = \lambda$, and work directly with the function $t_s(z)$. With this choice of parameterization, solving Eq.~\eqref{t_eq} with $f(z) = 1$ for $d>2$ results in two branches of solutions,
\begin{equation}\label{higher-d_vacuum_solution}
\begin{split}
&~~~~~~~t_\pm(z) = A_\pm \pm i \frac{z_t}{d}\left(\frac{z}{z_t}\right)^d \times \\
&{}_2F_1\left(\frac{1}{2},\frac{d}{2(d-1)}, \frac{3d-2}{2(d-1)}, \left(\frac{z}{z_t}\right)^{2d-2} \right), 
\end{split}
\end{equation}
where we have demanded analyticity at $z=0$. $z_t$ is the tip of extremal surface, and $A_\pm$ are integration constants. To fix these three quantities, we impose that the lower (upper) branch $t_-(z)$ ($t_+(z)$) ends at the lower (upper) boundary of the timelike strip $\mathcal{T}$, $t_\pm(z=\epsilon)=\pm \Delta t/2$, and that both branches meet continuously at the tip, $t_+(z_t) = t_-(z_t)$. At leading order in $\epsilon$,\footnote{There is an alternative choice where we identify $t_\pm(z=\epsilon)= \mp \Delta t/2$, which leads to a complex-conjugated $z_t$, and a complex-conjugated $S_\mathcal{T}$.} 
\begin{equation}\label{vacuum_solution_higher-d}
A_\pm = \pm \frac{\Delta t}{2}, \quad z_t =  \frac{i\Gamma\left(\frac{1}{2(d-1)}\right) }{2\sqrt{\pi} \Gamma\left(\frac{d}{2(d-1)}\right)}\Delta t.    
\end{equation}
A path in the complex $z$-plane, which starts at $z=\epsilon$ on the lower branch, goes from the lower to the upper branch at $z_t$, and finally ends at $z=\epsilon$ on the upper branch, provides a valid section of this complex extremal surface.  Evaluating the area density functional \eqref{action} along this path leads directly to Eq.~\eqref{ST_vacuum_higher-d} upon application of Eq.~\eqref{proposal}. 

\vspace{10 pt}

\noindent \emph{\textbf{Predictions for excited states}.--} The key aspect of the proposal~\eqref{proposal} is that it allows us to study HTEE when there are no other means to obtain it. As an important test bed we consider thermal states in CFT$_{3}$ on Minkowski space, represented holographically by $d=3$ black branes. The main novelty with respect to the cases considered so far is the existence of several distinct complex extremal surfaces $\gamma_T$ associated to the same boundary region $\mathcal{T}$.

\begin{figure}[h!]
\begin{center}
\includegraphics[width=\linewidth]{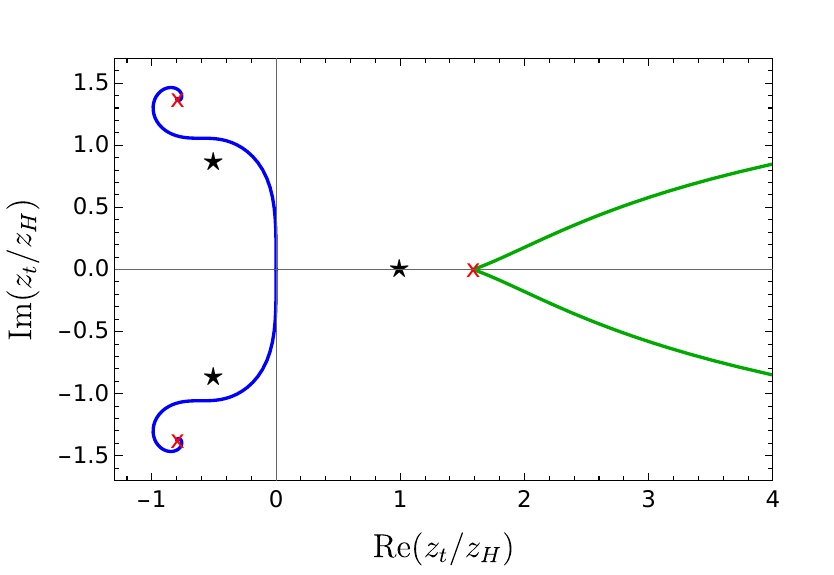}
\caption{\small $z_t$ for all the known complex extremal hypersurfaces in an AdS$_4$-Schwarzschild black brane. Blue (green) curves correspond to v.c. (v.d.) solutions. Horizons [roots of $f(z) = 0$] are represented as black stars, and critical extremal surfaces  as red crosses.} 
\label{fig:d=3_solution_space}
\end{center}
\end{figure}

Specifically, for a given $\Delta t \geq 0$, there exist two different classes of complex extremal surfaces to consider. Since Eq.~\eqref{t_eq} has real coefficients, each class comprises two branches of complex extremal surfaces related by a complex conjugation. We refer to these two classes as the \emph{vacuum-connected} (v.c.) and the \emph{vacuum-disconnected} (v.d.) solutions (see Fig.~\ref{fig:d=3_solution_space} for the location of their tips $z_t$ in the complex $z$-plane). The reason behind this nomenclature is as follows. In the $\Delta t \to 0$ limit, the tips of the v.c. solutions approach the asymptotic boundary at $z=0$ and, at leading order in $\Delta t$, are given by Eq.~\eqref{vacuum_solution_higher-d} (for the upper $\Im(z_t) \geq 0$ branch) or its complex conjugate (for the lower $\Im(z_t) \leq 0$ branch). On the other hand, in the same limit, the tips of the v.d. solutions approach the black brane singularity at $z=\infty$. The behavior of the conserved momentum $p$ also differs between the two classes of solutions: as $\Delta t\to 0$, $|p|$ diverges in the v.c. case, while it goes to zero in the v.d. one.  

To understand the behavior of both solution classes in the opposite, $\Delta t \to \infty$ regime, we need to recall the notion of a \emph{critical extremal surface} \cite{Hartman:2013qma}. A critical extremal surface is a solution of the equations of motion such that $z_s(\lambda) = z_c$. The location of the critical extremal surface in the complex $z$-plane is set by the requirement that the Lagrangian \eqref{action} evaluated on the critical extremal surface is stationary with respect to $z_c$, 
\begin{equation}\label{critical_point_eq}
\partial_{z_c} \sqrt{-\frac{f(z_c)}{z_c^{2d-2}}} = 0. \end{equation}
We will refer to $z_c$ as a critical point. In the case at hand, $f(z) = 1-\frac{z^3}{z_H^3}$, and Eq.~\eqref{critical_point_eq} reduces to $z_c^3 = 4z_H^3$, with critical points $z_1 = 2^\frac{2}{3}z_H$, $z_2 = 2^\frac{2}{3} e^{\frac{2 \pi i}{3}}z_H$, and $z_3 = 2^\frac{2}{3} e^{-\frac{2 \pi i}{3}}z_H$. While these critical extremal surfaces do not satisfy the boundary condition $\partial \gamma_\mathcal{T} = \mathcal{T}$, they do govern the behavior of the valid solutions in the $\Delta t \to \infty$ limit. In this regime, the tips of both branches of v.d. solutions approach $z_1$, while the tips of the upper (lower) branch of v.c. solutions approach $z_2$ ($z_3$).  

\begin{figure}[h!]
\begin{center}
\includegraphics[width=\linewidth]{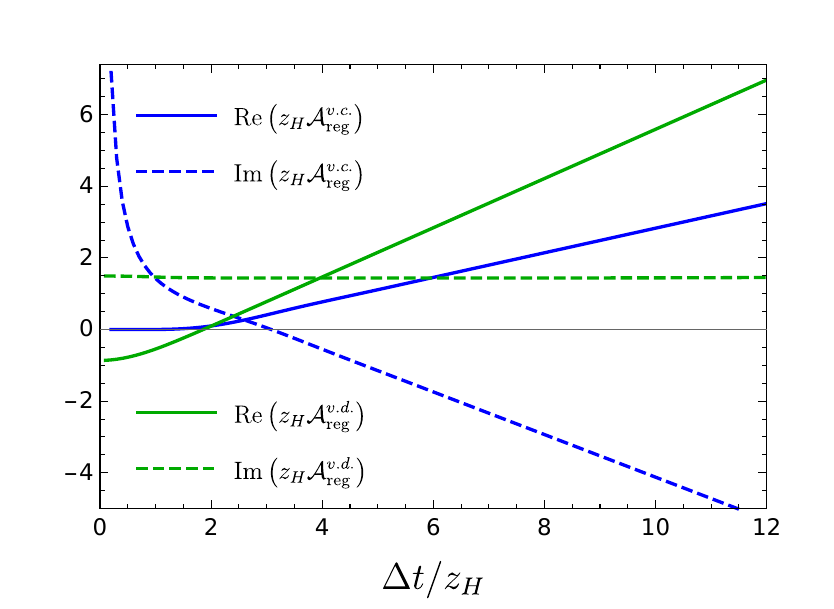}
\caption{\small Regularized area density $\mathcal{A}_\textrm{reg}$ for the v.c. (blue curves) and v.d. (green curves) extremal surfaces. Real (imaginary) parts correspond to solid (dashed) curves.} 
\label{fig:A_d=3_M}
\end{center}
\end{figure}

We define the finite part of the area density $\mathcal{A}$, $\mathcal{A}_\textrm{reg}$, as
\begin{equation}
\mathcal{A}_\textrm{reg} = \lim_{\epsilon \to 0} \left(\mathcal{A} - \frac{2}{\epsilon}\right), 
\end{equation}
and use superscripts $v.c.$ and $v.d.$ to denote the v.c. and v.d. solutions. Fig.~\ref{fig:A_d=3_M} depicts $\mathcal{A}_\textrm{reg}^{v.c.}$ and $\mathcal{A}_\textrm{reg}^{v.d.}$ as functions of $\Delta t$. Here and in the following, we restrict to the branches with $\Im(z_t) \geq 0$ to avoid clutter [the complex-conjugated branches have equal $\Re(\mathcal{A}_\textrm{reg})$ and opposite $\Im(\mathcal{A}_\textrm{reg})$]. 

\begin{figure}[h!]
\begin{center}
\includegraphics[width=0.5\linewidth]{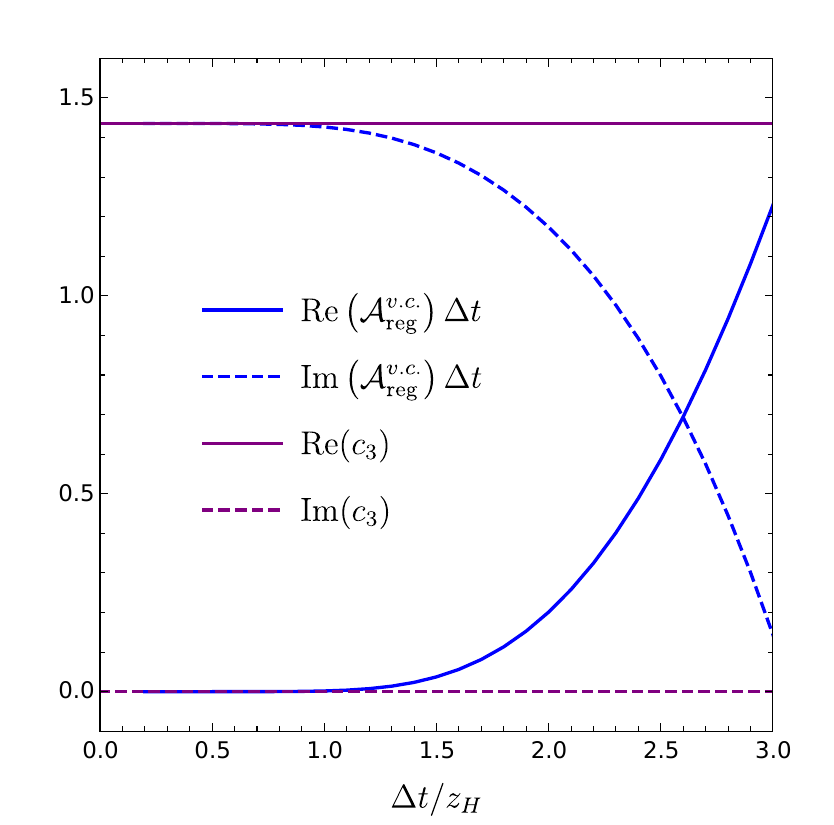}\includegraphics[width=0.5\linewidth]{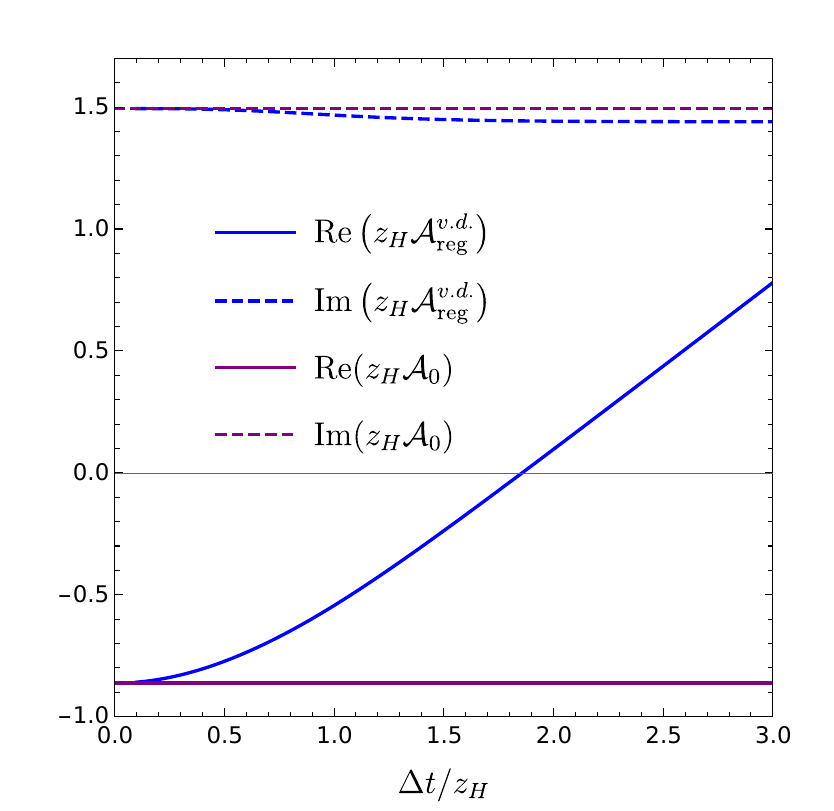}
\caption{\small Left: comparison between the $\Delta t \to 0$ limit of the regularized area density of the v.c. extremal surfaces, $\mathcal{A}_\textrm{reg}^{v.c.}$, and the vacuum result \eqref{A_vac}. Right: comparison between the $\Delta t \to 0$ limit of the regularized area density of the v.d. extremal surfaces, $\mathcal{A}_\textrm{reg}^{v.d.}$, and the prediction of the singularity probing solution, Eq.~\eqref{A_0}.} 
\label{fig:A_d=3_zoom-in_M}
\end{center}
\end{figure}

When $\Delta t \to \infty$, both $\mathcal{A}_\textrm{reg}^{v.c.}$ and $\mathcal{A}_\textrm{reg}^{v.d.}$ scale linearly with $\Delta t$, with a prefactor determined by their critical~points, 
\begin{equation}
\mathcal{A}_\textrm{reg}^{v.c.} \sim \frac{3^\frac{1}{2}}{2^\frac{4}{3}}e^{-\frac{i \pi}{3}} \frac{\Delta t}{z_H^2}, \quad  \mathcal{A}_\textrm{reg}^{v.d.} \sim \frac{3^\frac{1}{2}}{2^\frac{4}{3}}\frac{\Delta t}{z_H^2}. 
\end{equation} 
In the opposite, $\Delta t \to 0$ limit, the behavior of the finite part of both area densities is markedly different. For the v.c. solutions, $\mathcal{A}_\textrm{reg}^{v.c.}$ reduces to the vacuum result as expected (see Fig.~\ref{fig:A_d=3_zoom-in_M}), 
\begin{equation}\label{A_vac}
\mathcal{A}_\textrm{reg}^{v.c.} \sim i \frac{c_3}{\Delta t}. 
\end{equation}

On the other hand, in this regime, $\mathcal{A}_\textrm{reg}^{v.d.}$ does not exhibit power-law scaling with $\Delta t$, but rather goes to a constant. This constant has a straightforward geometric interpretation. Recall that, for the v.d. solutions, $|p|\to0$ as $\Delta t \to 0$. At $p=0$, Eq.~\eqref{t_eq} allows for the trivial solution $t_s(\lambda)=t_0 \in \mathbb R$, for which the area density functional reads 
\begin{equation}\label{A_vd_p=0}
\mathcal{A} = \int dz \left[z^4 \left(1-\frac{z^3}{z_H^3}\right)\right]^{-\frac{1}{2}}. \end{equation}
Evaluating Eq.~\eqref{A_vd_p=0} along a path in the complex $z$-plane that first goes from $z=\epsilon$ to $z=\infty$ slightly above the real axis, then crosses the branch cut at $z=\infty$, and finally comes back to $z=\epsilon$ slightly above the real axis again, results in
\begin{equation}\label{A_0}
\mathcal{A} \equiv \frac{2}{\epsilon} + \mathcal{A}_0 = \frac{2}{\epsilon} + \frac{2(-1+3^\frac{1}{2}i)\pi^\frac{1}{2} \Gamma\left(\frac{2}{3}\right)}{\Gamma\left(\frac{1}{6}\right)z_H}.    \end{equation}
In the right plot of Fig.~\ref{fig:A_d=3_zoom-in_M}, we compare $\mathcal{A}_0$ with the $\Delta t \to 0$ limit of $\mathcal{A}_\textrm{reg}^{v.d.}$, finding perfect agreement.\footnote{For the lower $\Im(z_t) \leq 0$ branch of v.d.~solutions, the relevant path in the complex $z$-plane goes below the positive real axis.} We emphasize that, while at $\Delta t = 0$ the v.d. solutions pierce the singularity, for any $\Delta t \neq 0$ they correspond to completely smooth complex extremal surfaces with a tip close to, but away from it.  

We will comment on the implication of the two classes of solutions for the computation of the HTEE in the outlook. 

\vspace{10 pt}

\noindent \textbf{\emph{Outlook}.--} Our Letter postulates that HTEE is defined in terms of complex extremal surfaces anchored in a timelike boundary subregion. Our explicit studies demonstrated that in general there are multiple complex extremal surfaces satisfying the same boundary conditions. This should not come as a surprise, given that an analogous phenomenon occurs for HEE, but it leaves us with the key question of which one computes the HTEE. 

We see two main possibilities to consider. The first one is to pick the surface with a minimal real part of the area, in analogy with HEE. In the black brane case we considered, this implies that the v.d. solutions dominate for small $\Delta t$ (see Fig.~\ref{fig:A_d=3_M}) and hence that the HTEE does not reduce to its vacuum counterpart as the temporal width of the boundary subregion tends to zero (see Fig.~\ref{fig:A_d=3_zoom-in_M}). For this choice, the HTEE violates the basic spirit of the UV--IR correspondence and is instead endowed with a UV--UV character, since the short-distance regime in the boundary CFT corresponds to the short-distance, sub-AdS length regime in the complexified bulk spacetime.

The second possibility is to regard the HTEE as the analytic continuation of the HEE when the boundary subregion is taken from spacelike to timelike in a specific manner. Note that, when the extent of the initial spacelike subregion tends to zero, the HEE reduces to the vacuum answer. Hence, while this analytic continuation is hard to implement, a basic requirement one can demand is that the same property holds for the HTEE of the final timelike subregion. This way of proceeding restricts the relevant surfaces in the HTEE computation and restores the UV--IR correspondence.

We emphasize that, irrespectively of the chosen discrimination criterion, it is perfectly possible that whichever complex extremal surface exists and does not happen to be contributing to HTEE has alternative physical interpretation, in analogy to the role played by entwinement for HEE \cite{Balasubramanian:2014sra, Craps:2022pke}. Studies of further examples will certainly be insightful in this respect. 

Finally, in a broader context, our work raises the question of whether there are contributions to various holographic observables that originate from complex extremal surfaces and were missed in the literature (a possibility that has not gone unnoticed \cite{Fischetti:2014zja,Fischetti:2014uxa}). This provides another arena where the methods developed in the present Letter apply.\\

\begin{acknowledgments}
We would like to thank J.~Harper, J.~Haegeman, M.~Mezei, R.~C.~Myers, L.~Tagliacozzo, T.~Takayanagi, W.~Tang and B.~Withers for discussions and comments on the draft. This project has received funding from the European Research Council (ERC) under the European Union’s Horizon 2020 research and innovation programme (grant number: 101089093 / project acronym: High-TheQ). Views and opinions expressed are however those of the authors only and do not necessarily reflect those of the European Union or the European Research Council. Neither the European Union nor the granting authority can be held responsible for them. 
\end{acknowledgments}

\bibliographystyle{bibstyl}
\bibliography{tlee} 

\clearpage
\onecolumngrid
\setcounter{page}{1}
\begin{center}
\textbf{\large Supplemental Material}
\vspace{1em}
\end{center}
\twocolumngrid

\appendix

\section{Sections of complex geodesics \\ vs. mixed-signature curves}
\label{app:ads3_sections}

Here we discuss the difference between sections of complex geodesics that can be considered within our HTEE proposal \eqref{proposal}, and the mixed-signature curves studied in Ref.~\cite{Doi:2023zaf}. To make direct comparison with the latter results, we will focus on the case of AdS$_3$/CFT$_2$ correspondence, both in the vacuum and thermal state.

\noindent \textbf{\emph{Vacuum.--}} The solution for a complex geodesic anchored at the boundary of a timelike interval in vacuum AdS$_3$ is given by Eq.~\eqref{ads3_sol1}. Since it is expressed in terms of the affine parameter $\lambda$, it describes an infinite set of complex geodesics connecting $\lambda=-\lambda_*$ to $\lambda=\lambda_*$, all with the same length $2\lambda_*$ \eqref{ads3_sol2}, corresponding to the infinite number of possible paths between the boundary points in the complex-$\lambda$ plane. However, specific choices of such paths lead to geodesics that, while still being complex, move in sections of the complexified spacetime with nontrivial reality conditions on the coordinates.

\begin{figure}
    \centering
    \includegraphics[width=\linewidth]{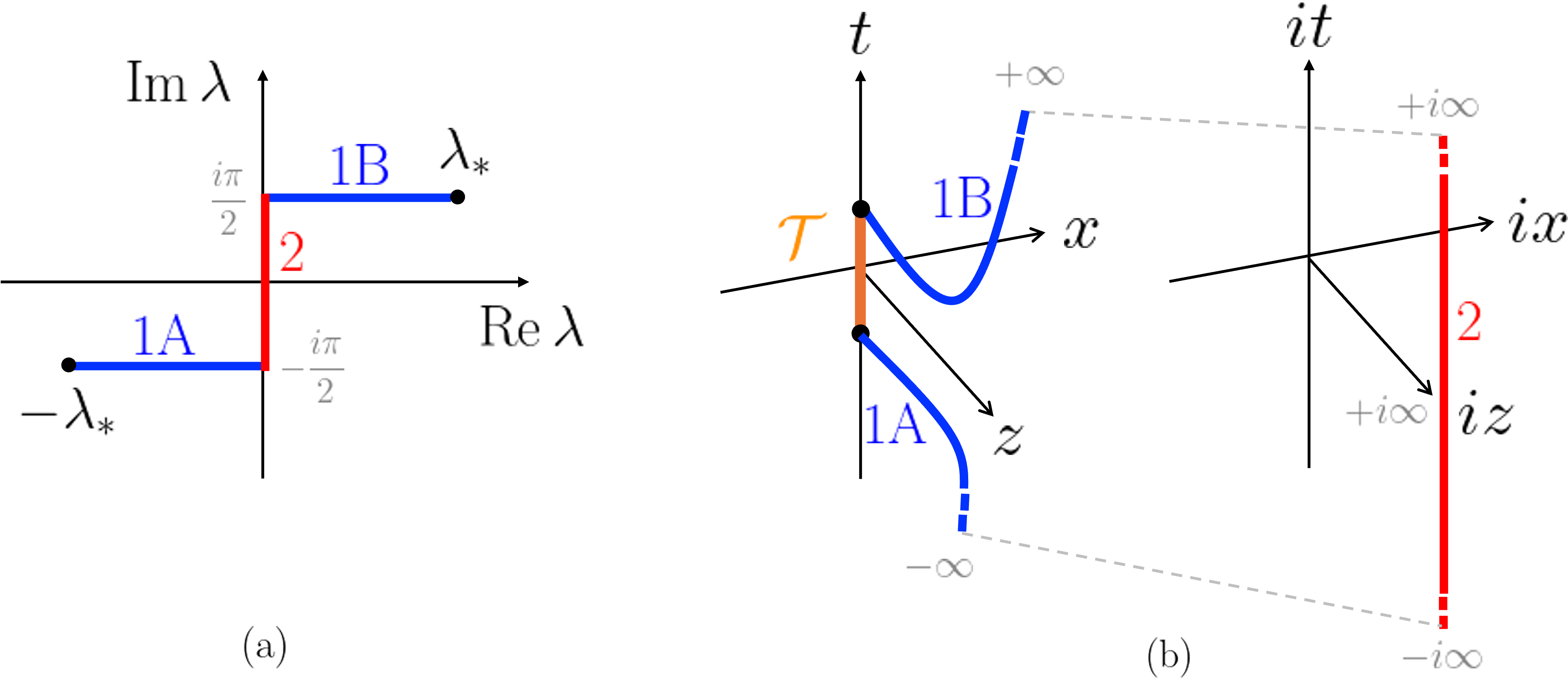}\vspace{2em}
    \includegraphics[width=.9\linewidth]{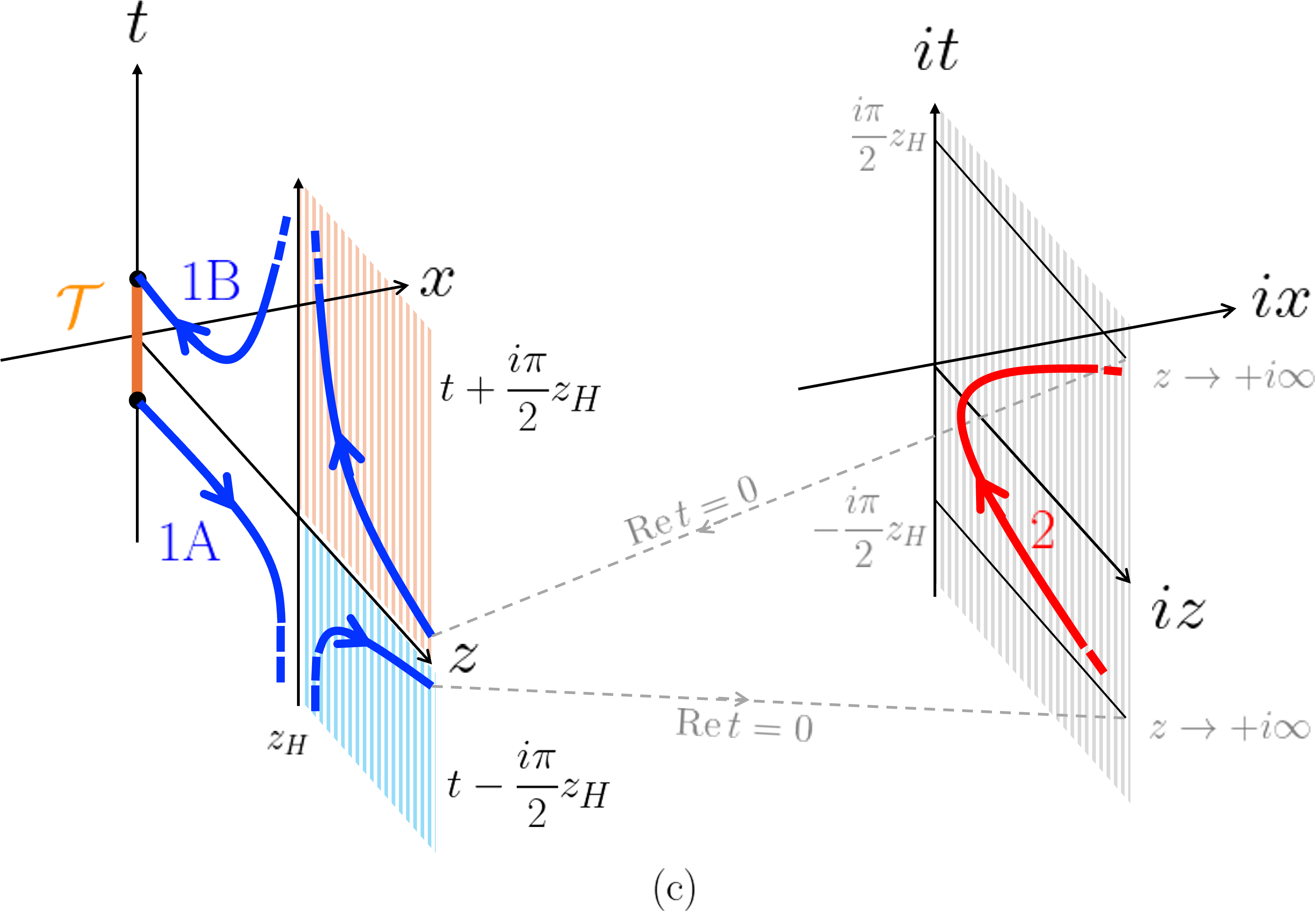}
    \caption{(a) A choice of path in the complex-$\lambda$ plane that resembles the most the piecewise curve considered in \cite{Doi:2023zaf} to describe HTEE in AdS$_3$. (b) The corresponding geodesic schematically depicted in vacuum AdS$_3$: note that the branches \textbf{1A}, \textbf{1B} live in a purely real spacetime, while \textbf{2} in a different section of the complex manifold, where all coordinates are purely imaginary. (c) The same as (b), for AdS$_3$ black brane with the horizon at $z=z_H$. After crossing the horizon, the two branches \textbf{1A}, \textbf{1B} acquire a constant imaginary part.}
    \label{fig:paths}
\end{figure}

The case which is interesting for our purposes is given by the path in Fig.~\ref{fig:paths}(a), which corresponds to moving:
\begin{description}
    \item[1A] from $-\lambda_*$ to $-i\pi/2$ at constant $\Im\,\lambda=-i\pi/2$;
    \item[2] from $-i\pi/2$ to $i\pi/2$ at constant $\Re\,\lambda=0$;
    \item[1B] from $i\pi/2$ to $\lambda_*$ at constant $\Im\,\lambda=i\pi/2$.
\end{description}

Along \textbf{1A} and \textbf{1B}, the coordinates $z(\lambda)$, $t(\lambda)$ are purely real, and evolve from $z(\pm\lambda_*)=\epsilon$ and $t(\pm\lambda_*)=\pm \Delta t/2$, to $z(\pm i\pi/2)=+\infty$, $t(\pm i\pi/2)=\pm \infty$. Hence, these branches of the path connect the cutoff surface, close to the boundary, to the Poincaré horizon, see Fig.~\ref{fig:paths}(b). Their length is real, and gives the real part of the entropy \eqref{ads3_vacuum_ent}. Note that these branches correspond exactly to the ones found in Ref.~\cite{Doi:2023zaf}, where they are however interpreted as piecewise spacelike curves. \textbf{1A} and \textbf{1B} are joined together by \textbf{2}, where instead both $z(\lambda)$ and $t(\lambda)$ are purely imaginary, and evolve from $z(-i\pi/2)=+i\infty$ and $t(-i\pi/2)=-i\infty$, to $z(i\pi/2)=+i\infty$ and $t(i\pi/2)=+i\infty$, see again Fig.~\ref{fig:paths}(b). Its length is purely imaginary, and gives the imaginary part of \eqref{ads3_vacuum_ent}. By letting $z=i\zeta$, $t=i\tau$ and $x=i\chi$, this path can be equivalently interpreted as a timelike geodesic that joins two points at infinite past and future on the Poincaré horizon and at constant $\chi=0$, in a planar AdS$_3$ spacetime with metric
\begin{equation}\label{AdS_im}
ds^2=\frac{-d\tau^2+d\zeta^2+d\chi^2}{\zeta^2}\,.
\end{equation}
This timelike geodesic would correspond to the one considered in \cite{Doi:2023zaf}, if not for the fact that it does not live in the original AdS$_3$ spacetime, where all coordinates are real, but in a different section of the complexified AdS$_3$ manifold, where the coordinates are purely imaginary. In other words, the piecewise, mixed-signature curve described by Ref.~\cite{Doi:2023zaf} is not an extremum of the Lorentzian functional~\eqref{action}. A simple way to acknowledge this is to look at the conserved momentum $p$, which in this case is given by
\begin{equation}
p=-\frac{t'(\lambda)}{z(\lambda)^2}\,.
\end{equation}
If we assume that the piecewise curve has to live in a real spacetime, we need to identify $\tau\to t$, $\zeta\to z$ and $\chi\to x$ in the metric \eqref{AdS_im}. However, as a consequence, $p\to-ip$ in switching from a spacelike to a timelike curve. This issue does not arise in the full complexified spacetime, where the coordinates, according to the path chosen, switch reality condition as well, preserving the conserved momentum $p=(\Delta t)^{-1}$. This is the sense in which the relevant geodesic exists only in a complexified version of the spacetime.

\noindent \textbf{\textit{Thermal state.--}} The same path shown in Fig.~\ref{fig:paths}(a) can be reinterpreted in the case of the AdS$_3$ black brane. The solutions are given by Eq.~\eqref{ads3_thermal_sol}, and show that now the branches \textbf{1A} and \textbf{1B} extend from the boundary $z(\pm \lambda_*)=\pm \epsilon$, $t(\pm \lambda_*) =\pm \Delta t/2$ to the horizon $z=z_H$, where the real part of $t(\lambda)$ diverges. In crossing the horizon, $t(z)$ also acquires a constant imaginary shift of $\pm i\beta/4=\pm i\pi z_H/2$, where $\beta=2\pi z_H$ is the inverse temperature of the BTZ black hole, and the sign is the same as the one of the real part. Then, the geodesic reaches $z(\pm i\pi/2)=+\infty$ at time $t(\pm i\pi/2)=\pm i \pi z_H/2$. Therefore, we can say that such a geodesic probes the singularity, with an expected imaginary shift in time, see Fig.~\ref{fig:paths}(c). The branch \textbf{2}, instead, ranges from $z(- i\pi /2)= +i\infty$, $t(-i\pi/2)=-i\pi z_H/2$ to $z(i\pi /2)= +i\infty$, $t(i\pi/2)=i\pi z_H/2$: in other words, it connects two points in the deep bulk of the spacetime, along the imaginary $z$-direction, through the imaginary $t$-direction, as shown in Fig.~\ref{fig:paths}(c).

Again, a comparison with Ref.~\cite{Doi:2023zaf} reveals a difference. The spacelike curves considered in \cite{Doi:2023zaf} correspond to our branches \textbf{1A} and \textbf{1B}, giving rise to the real part of the entropy \eqref{eq.HTEEbh3}. On the contrary, the timelike curve discussed in \cite{Doi:2023zaf}, which connects the past and future singularity by passing through the bifurcation point, is not realized as an extremum of the Lorentzian action principle in the real spacetime. The geodesic \textbf{2}, indeed, exists only in a different section of the complexified spacetime, where all the coordinates are imaginary, as it occurred for the vacuum state. We can still make the change of variables $z=i z_H\zeta$, $t=iz_H\tau$ and $x=iz_H\chi$ in the metric \eqref{AdS_im}, to get a geodesic now moving at constant $\chi=0$ in global AdS$_3$, with metric
\begin{equation}
ds^2=\frac{1}{\zeta^2}\left(-f(\zeta)d \tau^2+\frac{d\zeta^2}{f(\zeta)}+d\chi^2\right),
\end{equation}
where $f(\zeta)=1+\zeta^2$. Still, this section cannot be embedded naturally in the real one, since under the above change of coordinates the momentum $p$ switches reality condition as well. Again, the only possibility consistent with an extremization of the length functional is spacetime complexification.

\section{Mixed-signature hypersurfaces in AdS$_{d+1}$}
\label{app:adsd_sections}

It is natural to ask whether these mixed-signature piecewise curves could be the counterpart of higher-dimensional piecewise hypersurfaces in empty AdS$_{d+1}$, $d>2$. Despite not being extremal surfaces in the real spacetime, they can probe specific sections of the complexified one with special reality conditions on the coordinates. In this appendix we shall see that this is not the case in general, so that spacetime complexification is needed to describe HTEE across dimensions for timelike strips \eqref{strip}.

The fact that the complex extremal surfaces are anchored at the asymptotic boundary, regulated as usual as $z=\epsilon$, imposes that $t(z=\epsilon)\in\mathbb{R}$. If we consider the expression for $t(z)$ in the higher-dimensional case \eqref{higher-d_vacuum_solution}, and expand around $z=\epsilon\to 0$, we find
\begin{equation}
t_\pm(z) \mp \frac{\Delta t}{2} \sim \pm \alpha(iz)^d, \quad \alpha \in \mathbb R\,.
\end{equation}
For odd $d$, it is impossible to keep both $t$ and $z$ real as the hypersurface departs from the asymptotic boundary: they are in any case related by a factor of $i$. It is therefore impossible to construct a piecewise hypersurface in real coordinates as discussed in the previous appendix and in Ref.~\cite{Doi:2023zaf} for geodesics in AdS$_3$.

On the other hand, for even $d$ a piecewise real surface can be found in a similar way as described in Fig.~\ref{fig:paths}(b) for geodesics. The branches \textbf{1A} and \textbf{1B} correspond now to hypersurfaces anchored at $t(z=\epsilon)=\pm\Delta t/2$ and extending to infinity in both $t$ and $z$. Their area is precisely the cutoff-dependent term in \eqref{ST_vacuum_higher-d}. Branch \textbf{2}, instead, connects the two points at infinity along a direction where both $t$ and $z$ are purely imaginary and $z\to +i\infty$, see again Fig.~\ref{fig:paths}(b). The area of this branch can be understood by comparing the area density functional \eqref{action}, with $f(z)=1$, in two different parametrizations $z=z(t)$ and $t=t(z)$,
\begin{equation}\label{Adensities}
\mathcal{A}=\int_{-i\infty}^{+i\infty} dt\, \sqrt{\frac{1-z'(t)^{-2}}{z(t)^{2(d-1)}}}=2\int_{z_t}^{+i\infty}dz\, \sqrt{\frac{1-t'(z)^2}{z^{2(d-1)}}}
\end{equation}
where $z_t$ is the tip of the surface. In the limit of our interest $z\to +i\infty$ the integral does not vanish, as the integrand diverges with $t'(z)\to +i\infty$. This can be seen by expanding the on-shell area density functional in the right-hand side of \eqref{Adensities} around $z\to +i\infty$,
\begin{equation}
\mathcal{A}=A_0+\mathcal{O}\left(z^{-(d+1)}\right),
\end{equation}
where, for $d$ even, $A_0$ is a real constant. More precisely, it is given by
\begin{equation}
A_0=\frac{(-i)^d}{\left(d-2\right)\Delta t^{d-2}} \left(\frac{2 \sqrt{\pi }\,  \Gamma \left(\frac{d}{2 (d-1)}\right)}{\Gamma \left(\frac{1}{2 (d-1)}\right)}\right)^{d-1},
\end{equation}
i.e., the subleading term in \eqref{ST_vacuum_higher-d}.
Hence the HTEE is purely real and the cutoff-dependent term comes from the two branches \textbf{1A}, \textbf{1B} anchored at the boundary, while branch \textbf{2} contributes with the subleading real term in Eq.~\eqref{ST_vacuum_higher-d}. We stress again, however, that a comprehensive geometric interpretation can only be achieved by complexifying the coordinates.

\end{document}